\documentclass{desyproc}
\newcommand{\ci}[1]{\cite{#1}}
\newcommand{\bi}[1]{\bibitem{#1}}
\newcommand{\lab}[1]{\label{#1}}

\newcommand{\ba}{\begin{eqnarray}}
\newcommand{\ea}{\end{eqnarray}}
\newcommand{\beqs}{\begin{eqnarray}}
\newcommand{\eeqs}{\end{eqnarray}}

\begin{document}
\title{GPDs of the Nucleons \\ and Elastic Scattering at LHC Energies}

\author{{\slshape  O.V. Selyugin}\\[1ex]
BLTPH, JINR, Dubna, 141980, Russia }

\contribID{selyugin\_oleg}
\desyproc{DESY-PROC-2009-xx}
\acronym{EDS'09} 
\doi  

\maketitle

\begin{abstract}
  Taking into account the electromagnetic and gravitational form factors,
  calculated from a new set of $t $-dependent GPDs,
  a new model including the soft and hard pomerons is
  build. In the framework of this model the qualitative description of all
  existing experimental data at $\sqrt{s} \leq 52.8 $ GeV, including
  the Coulomb range and large momentum transfers, is obtained with only
  4 free parameters. Predictions for LHC energies are made.
\end{abstract}

\section{Introduction}
	The dynamics of strong interactions  finds its most
  complete representation in elastic scattering at small angles.
  Only in this region of interactions can we measure the basic properties that
  define the hadron structure: the total cross section,
  the slope of the diffraction peak and the parameter $\rho(s,t) $.
  Their values
  are connected on the one hand with the large-scale structure of hadrons and
  on the other hand with the first principles which lead to the
  theorems on the behavior of the scattering amplitudes at asymptotic
  energies \ci{mart,roy}.

   The definition of the structure of the high-energy elastic hadron-hadron
   scattering amplitude at small angles in the new superhigh energy range
   is a fundamental research problem.  In this
   kinematical domain PQCD cannot be directly operative; however,
   there are numerous results derived in the framework of axiomatic
   field theory that can guide us both at the theoretical and
   experimental levels.

   The hard pomeron which is obviously present in deeply inelastic
  processes with the large $\epsilon_2 \approx 0.4 $~\cite{lnd-hp}
  will lead to a strong decrease of the energy at which the saturation
  or black-disk regime appears. It is not obvious how the total cross sections
  will grow with energy, especially in the energy region of the LHC.
  In the present work I investigate the impact of the hard pomeron on
  some of the features of elastic proton-proton scattering at LHC
  energies and small momentum transfer.

   The situation is complicated by the possible transition to the saturation
regime, as the Black Disk Limit (BDL) will be reached at the LHC
\cite{SelyuginBDLCJ04,SelyuginBDL06}.
The effect of  saturation will be a change in the $t $-dependence of $B $ and $\rho $,
which will begin for $\sqrt{s} = 2 $ to  6  TeV, and which may drastically
change $B(t) $ and $\rho(t) $ at $\sqrt{s} = 14 $~TeV \cite{SelyuginBDL06,CSPL08}.
As we are about to see, such a feature can be obtained in very different models.

       There are indeed many different models for the description of hadron elastic
 scattering at small angles \cite{Rev-LHC}
They lead to  different
 predictions for the structure of the scattering amplitude at asymptotic
 energies, where the diffraction  processes can display complicated
 features \cite{dif04}.  This concerns especially the asymptotic unitarity
 bound connected with the Black Disk Limit (BDL) \cite{CPS-EPJ08}.

  The total helicity amplitudes can be written as $\Phi_{i}(s,t) =
  \Phi^{h}_{i}(s,t)+\Phi^{\rm em}_{i}(s,t) e^{\varphi(s,t)} $\,, where
 $\Phi^{h}_{i}(s,t) $ comes from the strong interactions,
 $\Phi^{\rm em}_{i}(s,t) $ from the electromagnetic interactions and
 $\varphi(s,t) $
 is the interference phase factor between the electromagnetic and strong
 interactions \cite{selmp1,selmp2,Selphase}.
 For the hadron part the amplitude with spin-flip is neglected, as usual at high energy.

  In practice,  many different partial waves with
 $l \rightarrow \infty $ must be summed and this leads to the impact parameter
 representation \cite{Predazzi66} converting
 the summation over $l $  into an integration over the impact parameter $b $.
 In the  impact-parameter representation
  the  Born term of the scattering amplitude will be
 \begin{eqnarray}
 \chi(s,b) \  \sim 
   \ \int \ d^2 q \ e^{i \vec{b} \cdot \vec{q} } \  F_{\rm Born}\left(s,q^2\right)\,,
 \label{tot02}
 \end{eqnarray}
 where $t= -q^2 $ and dropping the kinematical factor $1/\sqrt{s(s-2m_p^2)} $
 and a factor $s $ in front of the scattering amplitude.
  After  unitarisation, the scattering amplitude becomes
  \begin{eqnarray}
 F(s,t)  \sim
    \ \int \ e^{i \vec{b} \vec{q} }  \ \Gamma(s,b)   \ d^2 b\,.
 \label{overlap}
 \end{eqnarray}
 The overlap function $\Gamma(s,b) $ can be a matrix,
 corresponding to the scattering of different spin states.
 Unitarity of the $S $-matrix, $SS^{+} \leq 1 $, requires that $\Gamma(s,b) \leq 1 $.
 There can be different unitarization schemes which map $\chi(s,b) $ to different regions
 of the unitarity circle \cite{CSPPRD09}.
In this work I used the standard eikonal unitarisation scheme which leads to the
standard regime of saturation, {\it i.e.} the  BDL \cite{Bog100}:
     \begin{eqnarray}
  \Gamma(s,b)  = 1- \exp[-i\chi(s,b)] .
 \label{overlap}
\end{eqnarray}

\section{Born amplitude in the impact-parameter representation}
 In different models  one can obtain various pictures of
the profile function based on different representations of the hadron structure.
     In this model I suppose that the elastic hadron scattering amplitude can be divided in
   two pieces. One is proportional to the electromagnetic form factor. It plays the most important
    role at small momentum transfer. The other piece is proportional to
    the matter distribution in the hadron
    and is dominant at large momentum transfer.

As in the EPSH model~\cite{CSPL08}, I take into account the contributions of the soft and hard pomerons.
In this approach the nucleon-nucleon  elastic scattering amplitude is
proportional to the electromagnetic hadrons form-factors and can be approximated
at small $t $  by
\begin{eqnarray}
 T(s,t) \ =  [\ k_{1} \ (s/s_0)^{\epsilon_1}
           e^{\alpha^{\prime}_1 \  t \ ln (s/s_0)}
         \ + \    \ k_{2} \ (s/s_0)^{\epsilon_2}
             e^{\alpha^{\prime}_2 \  t \ ln (s/s_0)} ]
   \ G_{em}^2(t),
\end{eqnarray}
where $k_1=4.47 $  and $k_2 = 0.005 $ are the coupling of the ``soft''
 and ``hard'' pomerons, and $\epsilon_1 =0.00728 $, $\alpha^{\prime}_1=0.3 $,
 and $\epsilon_2=0.45 $, $\alpha^{\prime}_2=0.10 $
 are the intercepts and the slopes of the two pomeron trajectories.
The normalization $s_0 $ will be dropped below and
 $s $ contains implicitly the phase factor $\exp(-i \pi/2) $.
  I shall examine only high-energy nucleon-nucleon scattering with
 $\sqrt{s} \geq 52.8 $~GeV. So, the contributions
  of reggeons and odderon will be neglected. This model  only includes crossing-symmetric scattering
  amplitudes. Hence the differential cross sections of the proton-proton and proton-antiproton
  elastic scattering are equal.

The assumption about the hadron form-factors leads to the amplitude
      \begin{eqnarray}
    T(s,t)_{Born.} = h_1 (F^{s}_{Born} +
    F^{h}_{Born})G_{em}^2+ h_2 (F^{s}_{Born} + F^{h}_{Born})G_{grav.}^2,
    \end{eqnarray}
  I suppose a non-linear trajectory for the pomeron and, as a first approximation, assume
  that the coupling is proportional to the gravitational form factor and that
  both soft and hard terms in the $F_{Born}(s,t) $ have $\alpha^{'}=0 $ at large $t $.

\section{Hadron form factors}
  As was mentioned above,  all the form factors are obtained from the GPDs of the nucleon.
The electromagnetic form factors can be represented as first  moments of GPDs
\ba
 F_{1}^q (t) = \int^{1}_{0} \ dx  \ {\cal{ H}}^{q} (x, t); \ \ \
 F_{2}^q (t) = \int^{1}_{0} \ dx \  {\cal{E}}^{q} (x,  t),
\ea
following from the sum rules \cite{Ji97,R97}.

 Recently, there were many different proposals for the $t $ dependence of GPDs.
 We introduced a simple form for this
 $t $-dependence~\cite{STGPD}, based on the original Gaussian form corresponding to that
 of the wave function of the hadron. It satisfies the conditions of non-factorization,
 introduced by Radyushkin, and the Burkhardt condition on the power of $(1-x)^n $
 in the exponential form of the $t $-dependence. With this simple form
  we obtained a good description of the proton electromagnetic Sachs form factors.
  Using the isotopic invariance we obtained good descriptions of the neutron
  Sachs form factors without changing any  parameters.

 The Dirac elastic form factor can be written
\begin{eqnarray}
 G^2(t)= h_{fa} e^{d_1 \ t} \ + \  h_{fb} e^{d_2 \ t} \
 + h_{fc} e^{d_3 \ t}.   \lab{eff}
 \end{eqnarray}
with $h_{fa}=0.55 $, $h_{fb}=0.25 $, $h_{fa}=0.20 $, and
 $d_1=5.5 $, $d_2=4.1 $, $d_3=1.2 $.
    The exponential form of the form factor lets us calculate the hadron scattering amplitude
    in the impact parameter representation~\cite{CSPL08}.

I shall use this model of GPDs to obtain the gravitational form factor of the nucleon in the
impact-parameter representation.
This form factor can be obtained from the second momentum of the GPDs. Taking instead of
the electromagnetic current $J^{\mu} $ the energy-momentum tensor $T_{\mu \nu} $ together
with a model of quark GPDs, one can obtain the gravitational form factor of fermions
\ba
\int^{1}_{-1} \ dx \ x [H(x,\Delta^2,\xi) \pm E(x,\Delta^2,\xi)]  = A_q(\Delta^2) \pm B_{q}(\Delta^2) .
\ea
 For $\xi=0 $ one has
\ba
\int^{1}_{0} \ dx \ x [{\cal{H}}(x,t) \pm {\cal{E}}(x,t)] = A_{q}(t) \pm B_{q}(t).
\ea
Calculations in the momentum-transfer representation show
that the second moment of the GPDs, corresponding to the gravitional form-factor, can be
represented in the dipole form
\begin{eqnarray}
A(t)=L^2/(1-t/L^2)^2  .\label{overlap}
 \end{eqnarray}
with the parameter $L^2=1.8 $~GeV $^2 $.
For the scattering amplitude, this leads to
\begin{eqnarray}
A(s,b)  = \frac{L^5 b^3}{48} K_{3}(Lb), \label{K3}
 \end{eqnarray}
where $K_{3}(Lb) $ is the MacDonald function of the third order.
To match both parts of the scattering amplitude,  the
second part is multiplied by a smooth correction function which depends on the impact parameter
\begin{eqnarray}
\psi(b) = (1+\sqrt{r_{1}^2+b^2}/\sqrt{r_{2}^{2}+b^2}).\label{K3}
 \end{eqnarray}

\section{Description of the differential cross sections}
The model has only four free parameters, which are obtained from a fit to
the experimental data:
 $$h_1=1.09 \pm 0.004; \ \ \ h_2=1.57 \pm 0.006; r_{1}^2=1.57 \pm 0.02; \ \ \ r_{2}= 5.56 \pm 0.06. $$
I used all the existing experimental data at $\sqrt{s} \geq 52.8 $~GeV,
including the whole Coulomb region and up to the largest momentum transfers experimentally accessible.
In the fitting procedure,  only statistical errors were taken into account.
The systematic errors were used as an additional common normalization of the experimental data
from a given experiment. As a result, one obtains $\sum \chi^2_i /N \simeq 3. $ where $N=924 $ is
the number of experimental points. Of course, if one sums the systematic and
statistical errors, the $\chi^2/N $ decreases, to $2 $.
Note that the parameters are energy-independent.
The energy dependence of the scattering amplitude is determined
only by the intercepts of the soft and hard pomerons.

\label{sec:figures}
\begin{figure}
\includegraphics[width=0.5\textwidth] {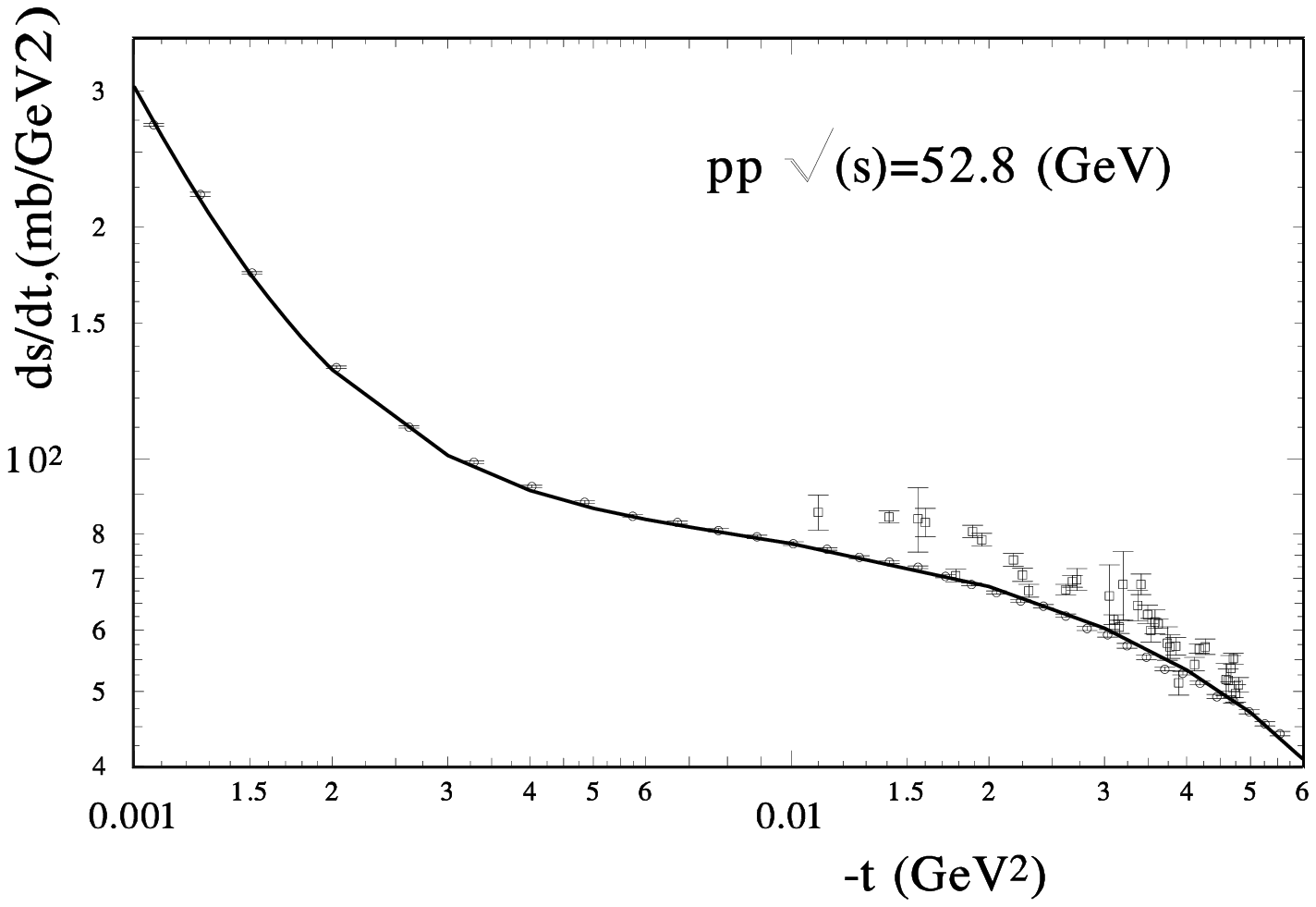}
\includegraphics[width=0.5\textwidth] {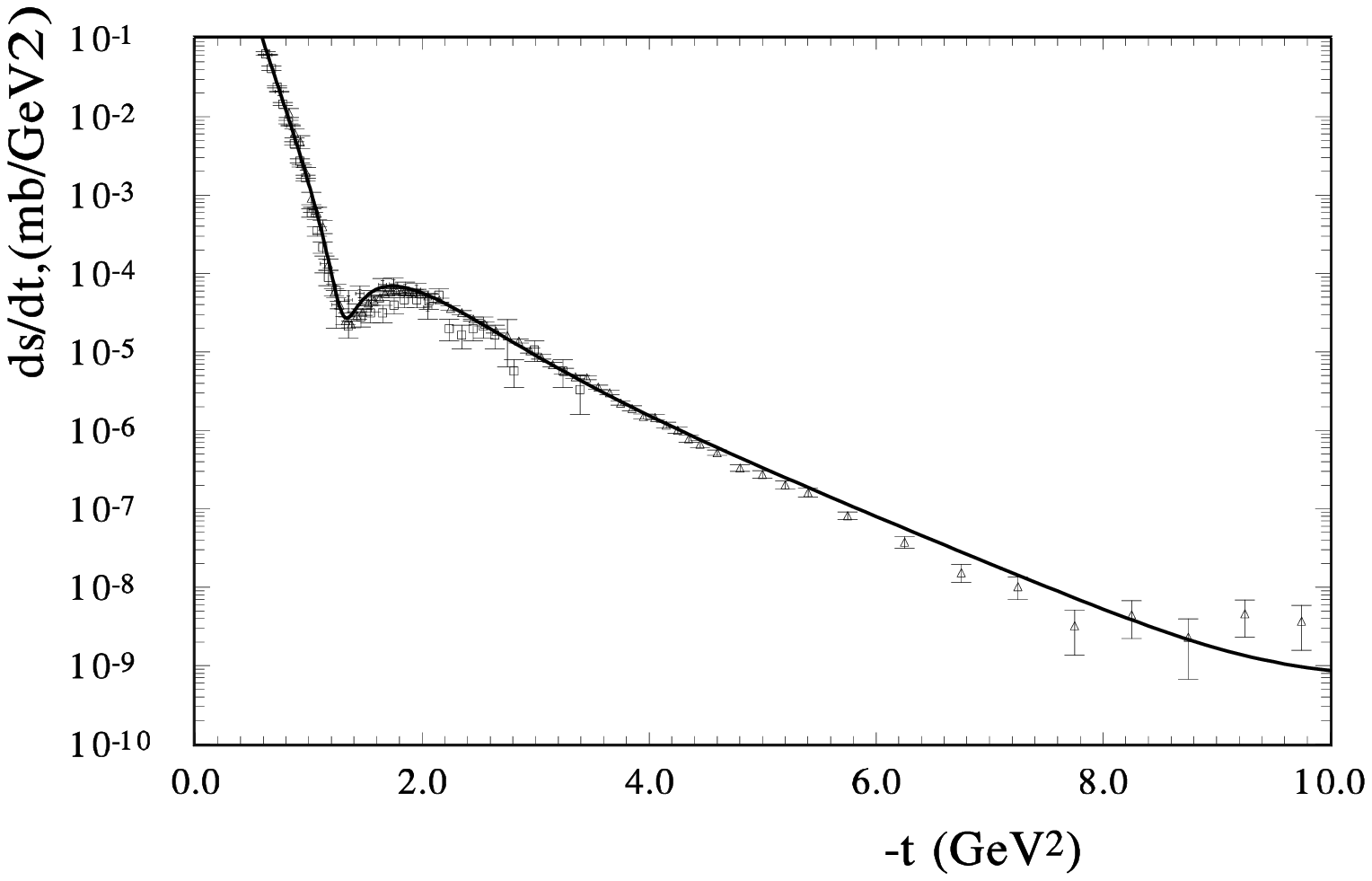}
\caption{ $d\sigma/dt \ {\rm at} \  \sqrt{s}=52.8 $~GeV for $pp $ elastic scattering,
at small $|t| $ (left) and at large $|t| $ (right). }\label{Fig:MV}
\end{figure}

In Fig.~1 the differential cross sections for proton-proton elastic scattering at
 $\sqrt{s} = 52.8 $~GeV
are presented. At this energy there are experimental data at small
(beginning at $-t=0.0004 $~GeV $^2 $) and large (up to $-t=10 $~GeV $^2 $) momentum transfers.
The model reproduces both regions and provides a qualitative description of the dip region
at $-t \approx 1.4 $~GeV $^2 $, for $\sqrt{s}=53 $~GeV $^2 $ and for $\sqrt{s}=62.1 $~GeV $^2 $,
 although only the low- $t $ abd high- $t $ regions have been fitted to.

Now let us examine the proton-antiproton differential cross sections.
In this case at low momentum transfer
the Coulomb-hadron interference term plays an important role and has the opposite sign.
The model describes the experimental data well. In this case, the first part of the scattering
amplitude determines the differential cross sections, and is dominated by the exchange of the soft
pomeron.
\label{sec:figures}
\begin{figure}
\includegraphics[width=0.5\textwidth] {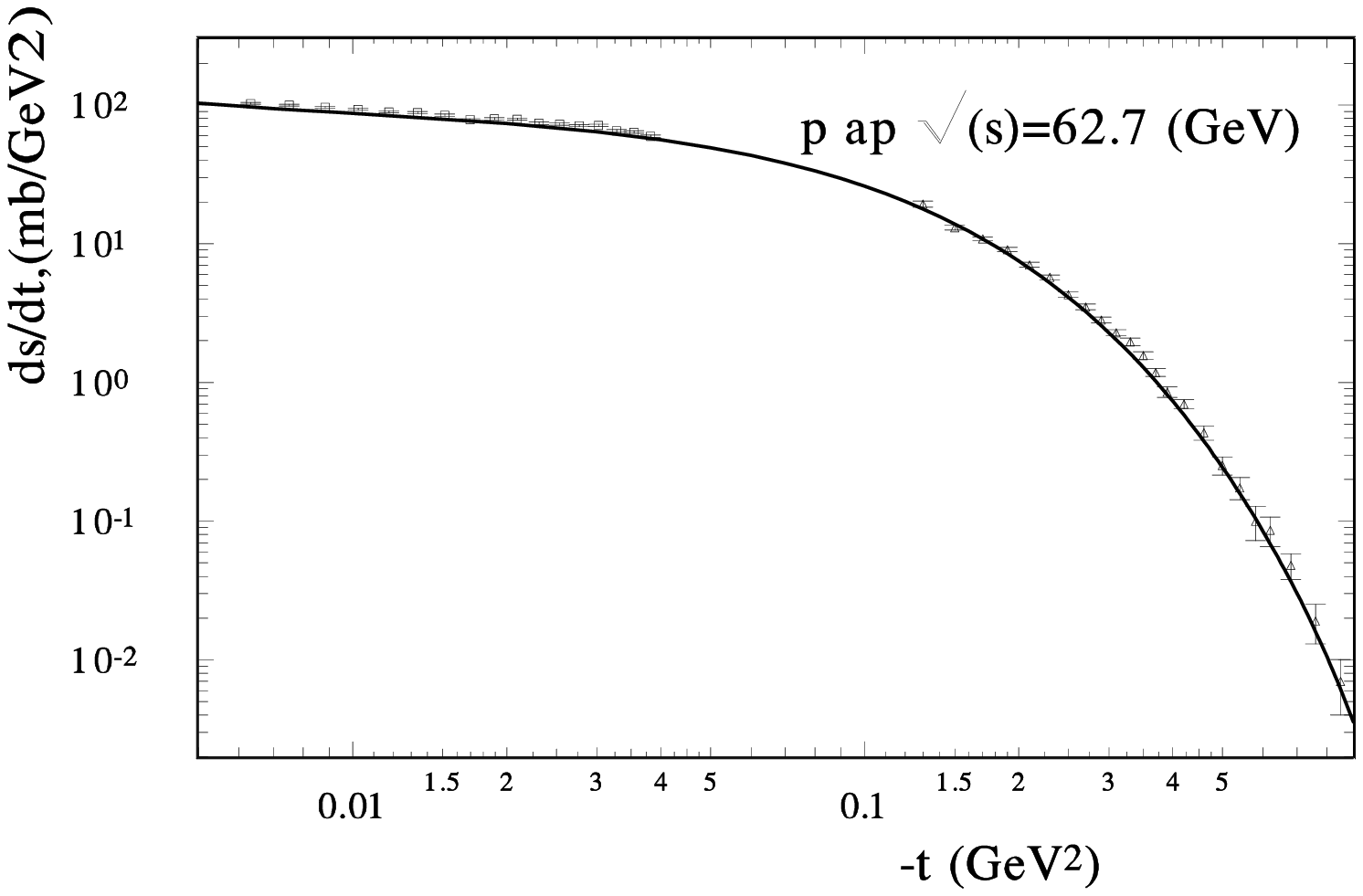}
\includegraphics[width=0.5\textwidth] {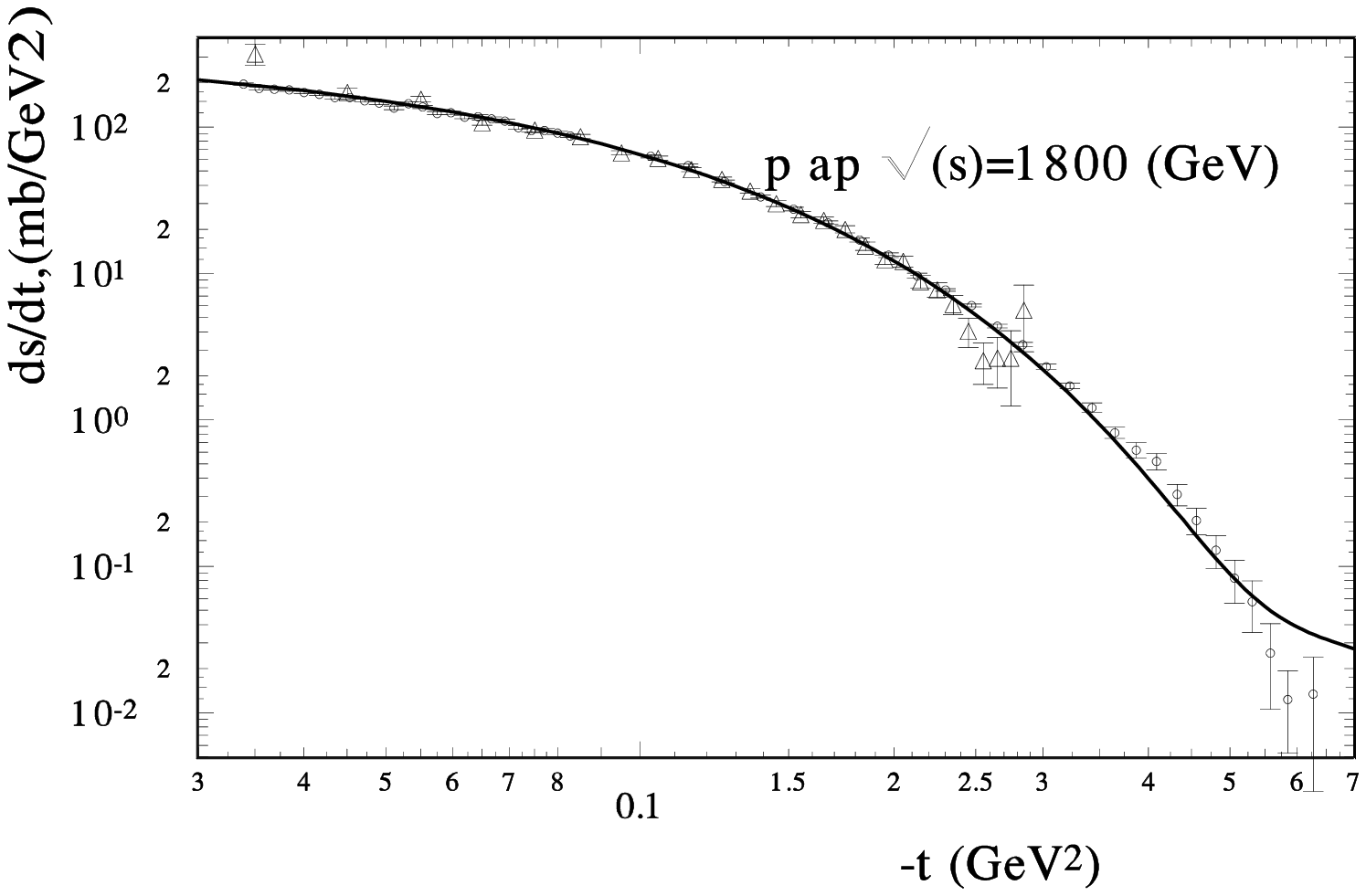}
\caption{ $d\sigma/dt $ for $\bar p p $ elastic scattering  at small $|t| $, at $ \sqrt{s}=541 $~GeV (left)
and $\sqrt{s}=1800 $~GeV (right)}\label{Fig:4}
\end{figure}

There was a significant difference between the experimental measurement of
 $\rho $, the ratio of the real part to the imaginary part of the scattering amplitude,
between the UA4 and UA4/2 collaborations at $\sqrt{s}=541 $~GeV.
As shown in Table 1, the resulting values for $\rho(0) $ appear inconsistent.
A more careful analysis~\cite{SelyuginYF92,SelyuginPL94}  shows
that there is no contradiction between the measurements of UA4 and UA4/2.
Now the present model gives for this energy $\rho(\sqrt{s}=541 {\rm GeV}, t=0) = 0.163 $,
so, practically the same as in the previous phenomenological analysis.
\label{sec:figures}
\begin{figure}
\includegraphics[width=0.5\textwidth] {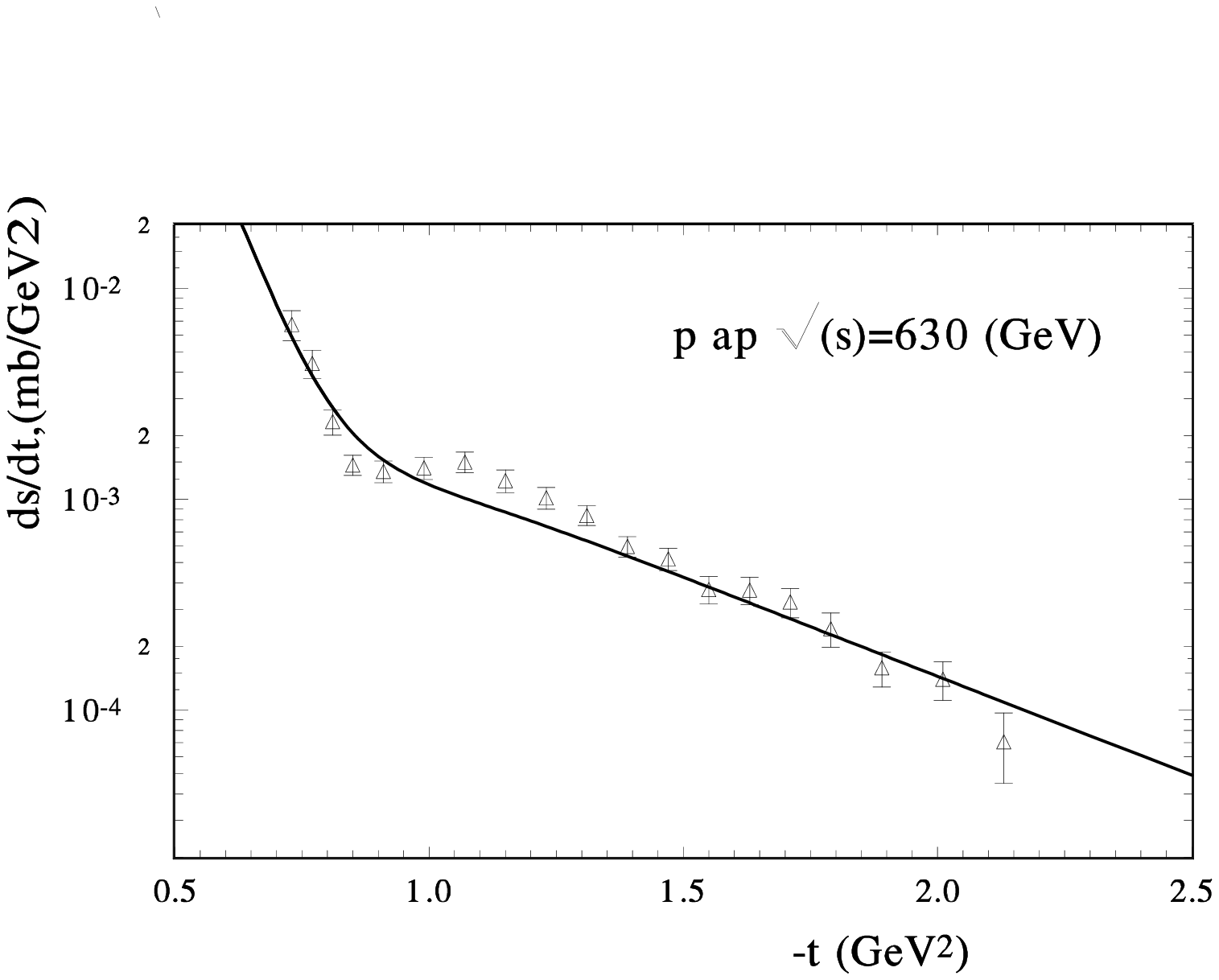}
\includegraphics[width=0.5\textwidth] {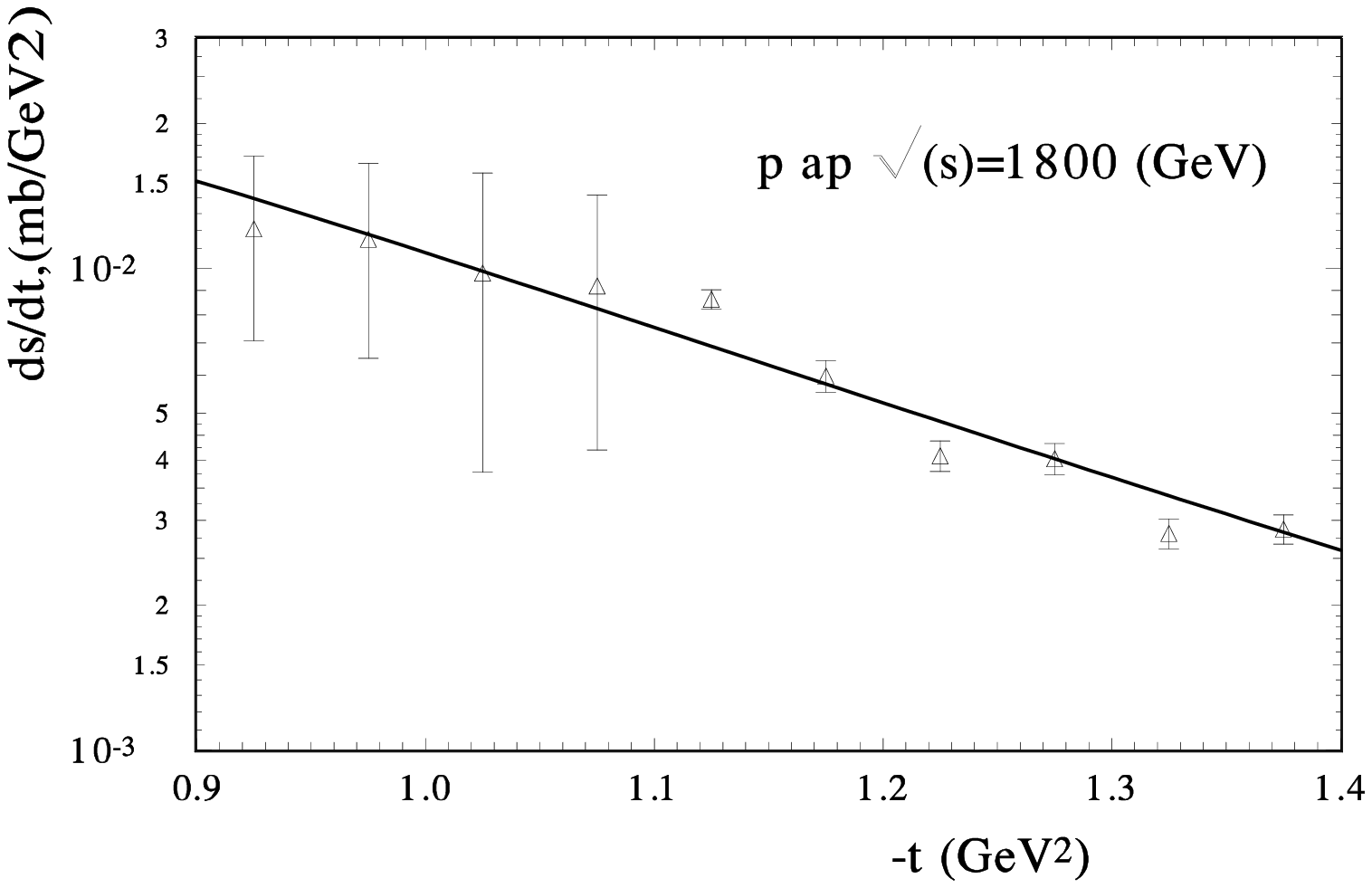}
\caption{ $d\sigma/dt $ for $\bar p p $ elastic scattering at large momentum transfer, at
 $\sqrt{s}=630 $~GeV (left) and
 $\sqrt{s}=1800 $~GeV (right)}\label{Fig:MV}
\end{figure}

\begin{table}[h]
\label{tab:2}       
\begin{center}
\begin{tabular}{ll|l|l}
\hline\noalign{\smallskip}
 \multicolumn{4}{c} { $\bar{\rho}  \ (\sqrt{s} =541 $~GeV, $ t=0) $ } \\
\noalign{\smallskip}\hline\noalign{\smallskip}
 experiment      & experimental analysis      & global analysis  & This model  \\
 UA4      & $0.24 \pm 0.02 $      & $0.19 \pm 0.03 $ [\cite{SelyuginYF92}]&  \\
 UA4/2    & $0.135 \pm 0.015 $     & $0.17 \pm 0.02 $  [\cite{SelyuginPL94}]&  0.163  \\
\noalign{\smallskip}\hline
\end{tabular}
\caption{Average values of $\rho $, derived with fixed total cross section (first two columns), and from a global analysis (last two columns).}
\end{center}
\end{table}

Now let us examine the data at higher energy, where the contribution of the hard pomeron
will be more important.
In Fig.~2 the description of the proton-antiproton scattering at $\sqrt{s}=541 $~GeV $^2 $ and at
 $\sqrt{s}=1800 $~GeV $^2 $ is shown. In this case the Coulomb-hadron interference term leads to a
large value of the real part of the scattering
amplitude, which is determined by the contribution from the hard pomeron.
The good description of the experimental data shows that the
 parameters of the hard pomeron correspond to the real physical situation.

Figure 3 shows the description of the experimental data at larger momentum transfers for
 $\sqrt{s}=630 $~GeV $^2 $ and $\sqrt{s}=1800 $~GeV $^2 $.  It is clear that the model leads to a
good description of these data. However, one must note that the fine structure of the dip
is not reproduced by the model in this case. The model shows only an essential change of the slope in
this region.

Saturation of the profile function will surely control the behavior of $\sigma_{tot} $ at higher
energies and will result in a significant decrease of the LHC cross section.
For the last LHC energy $ \sqrt{s}=14 $~TeV the model predicts
 $\sigma_{tot}=146 $~mb and $\rho(0)=0.235 $. This result comes from the contribution of the hard pomeron
 and from the strong saturation from the black disk limit.

\section{Conclusion}
In the presence of a hard Pomeron~\cite{clms2}, saturation effects can change the behavior of the
cross sections already at LHC energies. A new model, taking into account the contributions of the
soft and hard pomerons and using form factors calculated from the GPDs, successfully describes all
the existing experimental data on elastic proton-proton and proton-antiproton
scattering at  $\sqrt{s} \geq 52.8 $~GeV, including the Coulomb-hadron interference region,
the dip region, and the large-momentum-transfer region. The behavior of the differential cross
section at small  $t $ is determined by the electromagnetic form factors, and at large  $t $ by the
matter distribution (calculated in the model from the second momentum of the GPDs) as was supposed
by Mittinen a long time ago~\cite{Mittenen}.

The model leads to saturation of the BDL in the TeV region of energy.
As a result the parameters of the scattering amplitude
$B(s,t) $ and of $ \rho(s,t) $ have a complicated dependence on $s $ and $t $ and
the scattering amplitude has a non-exponential behavior at small momentum transfer.

The possibility of a new behavior of $\rho(s,t) $ and $B(s,t) $ at LHC energies~\cite{CSPL08}
has to be taken into account in the procedure extracting the value of the total cross sections
by the standard method ~\cite{CSPRL09}.

\section*{Acknowledgements}
I would like to thank for helpful discussions J.R. Cudell and O.V. Teryaev.
I gratefully acknowledges the organization committee and R. Orava  for the financial support
to take a part in the conference and
and would like to thank the  FRNS and University of Li\`{e}ge
where part of this work was done.

\begin{footnotesize}

\end{footnotesize}
\end{document}